\documentclass[aps,prb,reprint,superscriptaddress,showpacs]{revtex4-1}

\usepackage[dvipdfmx]{graphicx}
\usepackage{bm}
\usepackage{color}
\usepackage{amsmath,amssymb}

\begin{document}

\title{Odd-parity superconductivity in bilayer transition metal dichalcogenides}

\author{Yasuharu Nakamura}
\affiliation{Graduate School of Science and Technology, Niigata University, Niigata 950-2181, Japan}
\author{Youichi Yanase}
\email[]{yanase@scphys.kyoto-u.ac.jp}
\affiliation{Department of Physics, Graduate School of Science, Kyoto University, Kyoto 606-8502, Japan}

\date{\today}

\begin{abstract}
Spin-orbit coupling in transition metal dichalcogenides (TMDCs) causes spin-valley locking 
giving rise to unconventional optical, transport, and superconducting properties. 
In this paper, we propose exotic superconductivity in bilayer group-IV TMDCs by symmetry control. 
The sublattice-dependent ``hidden'' spin-orbit coupling arising from local inversion symmetry breaking 
in the crystal structure may stabilize the odd-parity superconductivity by purely $s$-wave local pairing interaction. 
The stability of the odd-parity superconducting state depends on the bilayer stacking. 
The 2H$_b$ stacking in Mo$X_2$ and W$X_2$ ($X$ =S, Se) favors the odd-parity superconductivity 
due to interlayer quantum interference. 
On the other hand, the odd-parity superconductivity is suppressed by the 2H$_a$ stacking of NbSe$_2$. 
Calculating the phase diagram of the tight-binding model derived from first principles 
band calculations, we conclude that the intercalated bilayer MoS$_2$ and WS$_2$ are candidates 
for a new class of odd-parity superconductors by spin-orbit coupling. 
\end{abstract}


\maketitle

\section{Introduction}

The recent fabrication of atomically thin transition metal dichalcogenides (TMDCs) films 
has led to extraordinary developments in both applied and basic 
sciences~\cite{Mak2010,Radisavljevic2011,Wang2012,Cao2012,Mak2012,Zeng2012,Zhang2014,Wu2014,Wu2015,Sangwan2015,
Mak2014,Song2013,Cui2015,Wu2013,Gong2013,Ye2012,Saito2015,Lu2015,Xi2015,Costanzo2016,Jo2015,Shi2015,Lu2017}.  
The discovery of superconductivity in MoS$_2$~\cite{Ye2012} and related TMDCs~\cite{Costanzo2016,Jo2015,Shi2015,Lu2017} 
has demonstrated a new paradigm of artificial two-dimensional (2D) superconductors.  
Indeed, an exotic superconducting phase 
protected by spin-orbit coupling has been identified~\cite{Saito2015,Lu2015,Xi2015,Lu2017}. 

\begin{figure}[htbp]
 \begin{center}
\includegraphics[keepaspectratio, width=9.0cm]{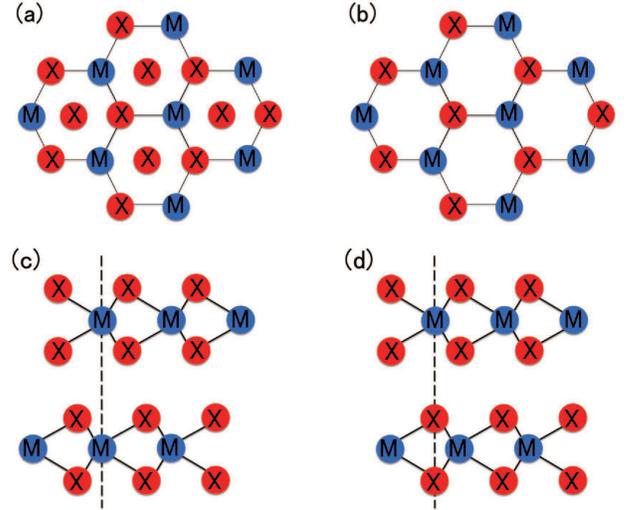}
  \caption{(Color online) 
Crystal structures of bilayer TMDCs. Blue M and red X show the metal ions and chalcogen ligands, respectively. 
(a) and (c) are top-view. (b) and (d) are side-view. 
(a) and (b) illustrate 2H$_a$ stacking structure, while (c) and (d) show 2H$_b$ stacking structure. 
}
  \label{fig:crystal_structure}
 \end{center}
\end{figure}

 The monolayer group-VI TMDCs have a hexagonal crystal structure sketched in Fig.~\ref{fig:crystal_structure}(b), 
where the metal ions are surrounded by six chalcogen ligands forming a prism structure~\cite{Xiao_review}. 
The point group symmetry is $D_{3h}$, lacking the inversion symmetry. Thus, the monolayer TMDCs are 
intrinsically noncentrosymmetric.  Therefore, the antisymmetric spin-orbit coupling (ASOC) 
appears and induces valley-dependent spin polarization along the crystallographic {\it c}-axis~\cite{Xiao2012}. 
Since the two valleys around the $K$ and $K'$ points in the Brillouin zone are time-reversal pairs, 
the ASOC induces such ``spin-valley locking''.  
The resulting spin-splitting of the band structure has been shown by first principles band structure 
calculations~\cite{Xiao_review,Zhu2011,Cheiwchanchamnangij2012,Kadantsev2012,Kosmider2013,Kormanyos2014,Coehoorn1987,Autieri2016} 
and was observed in many experiments~\cite{Xiao_review}.

A variety of intriguing phenomena caused by the spin-valley locking have been revealed by recent works. 
For instance, the superconducting state in the electron-doped MoS$_2$ is protected against the paramagnetic depairing effect. 
As a consequence, a huge upper critical field above $50$T, which significantly exceeds  
the Pauli-Chandrasekhar-Clogston limit~\cite{Chandrasekhar1962,Clogston1962}, 
has been observed~\cite{Saito2015}. Then, the superconductivity is called ``Ising superconductivity''~\cite{Lu2015,Xi2015}. 
A peculiar optical response~\cite{Xiao2012,Sanchez2013} has also been observed~\cite{Cao2012,Mak2012,Zeng2012,Zhang2014}.  
Furthermore, topological superconductivity in TMDCs~\cite{Yuan-Law2014,He-Law2016,Hsu2016} and 
TMDC-based devices~\cite{Zhou-Law2016,Zhang2016,Sharma2016,Wakatsuki-Law2016} has been theoretically proposed, and 
the possibility of the topological insulating phase was discussed~\cite{Cazalilla2014}.

\begin{table}[htbp]
\caption{Point group symmetry of monolayer, bilayer, and bulk TMDCs having 2H$_b$ structure.}
{\renewcommand\arraystretch{1.2}
\begin{tabular}{c|c|c|c}
& monolayer & bilayer & bulk \\ \hline
Point group & $D_{\rm 3h}$ & $D_{\rm 3d}$ & $D_{\rm 6h}$ 
\end{tabular}
} 
\label{tab0}
\end{table}
 Tunability of van der Waals heterostructures enables symmetry control in TMDCs from monolayer to bulk. 
The chemically stable stacking structure is determined mainly by the metal ions~\cite{Wilson1969}.
For examples, the 2H$_b$ structure is favored in group-VI Mo$X_2$ and W$X_2$ ($X$ =S, Se), while 
the 2H$_a$ structure is stable in group-V Nb$X_2$. 
The 2H$_a$ and 2H$_b$ stackings are sketched in Fig.~\ref{fig:crystal_structure}. 
In both structures, the 2D coordinates of metal ions and chalcogen ligands are exchanged 
between the upper and lower layers. 
Therefore, the space inversion symmetry is recovered when the number of TMDC layers is even. 
For the 2H$_b$ stacking, the point group is $D_{6h}$ in the bulk, and $D_{3d}$ in the bilayer, 
both of which preserve the space inversion symmetry. 
The symmetry of the heterostructures is summarized in Table~\ref{tab0}.

Although the global inversion symmetry is preserved in bilayer TMDCs, the local site symmetry of metal ions 
is still noncentrosymmetric $D_{3h}$.  
Such locally noncentrosymmetric crystal structure results in a sublattice-dependent ASOC~\cite{Maruyama2012,Fischer2011}. 
Although the spatial average of the ASOC disappears so as to preserve the global inversion symmetry, 
a sublattice-dependent spin polarization~\cite{Maruyama2012} has actually been 
observed in various materials~\cite{Goh2012,Shimozawa2014,Zhang2014-2,Riley2014,Jones2014,Ghelmann2016,Klein2016} including 
the TMDCs~\cite{Riley2014,Jones2014,Ghelmann2016}. 
As unusual superconducting properties have been established in noncentrosymmetric systems~\cite{NCSC,Agterberg_review}, 
it is naturally expected that unconventional superconducting states may be stabilized 
in locally noncentrosymmetric systems. 
Indeed, it has been shown that the ``spin-momentum-layer locking'' by the spin-orbit coupling may stabilize  
the odd-parity superconductivity in multilayer Rashba systems~\cite{Fu-Berg2010,Nakosai2012,Yoshida2012,Yoshida2014,Higashi2016} 
and a nonsymmorphic zigzag chain~\cite{Sumita2016}. Then, the spin-orbit coupling combined with the spin polarization 
causes the $\pi$-junction between the two sublattices and stabilizes the sign changing spin-singlet order parameter 
leading to odd-parity. Such a superconducting state has been identified as a pair-density-wave (PDW) state. 
Interestingly, multilayer odd-parity superconductivity is classified into 
the topological crystalline superconductivity~\cite{Yoshida2015,Watanabe2015,Yoshida_Tsuneya2016}, 
and zigzag chains are identified as $Z_2$ topological superconductors~\cite{Sumita2016}.
In this paper we propose the material realization of analogous odd-parity superconductivity in bilayer TMDCs.

 We show advantages of bilayer TMDCs for realizing the odd-parity superconductivity without 
spin-triplet pairing. 
First, in the 2H$_b$ structure, the interlayer hopping integral $t_\perp f(\bm k)$ vanishes at the $K$ and $K'$ point 
because of the quantum interference effect~\cite{Xiao_review,Akashi2014,Akashi2016}. 
Therefore, the ratio $\alpha/t_\perp f(\bm k)$, with $\alpha$ being the coupling constant of ASOC, is enhanced 
on the Fermi surface. This ratio is increased further by intercalation, which induces the superconductivity 
in K$_x$MoS$_2$, Rb$_x$MoS$_2$, and Cs$_x$MoS$_2$ without gating~\cite{Woollam1977}. 
The large $\alpha/t_\perp f(\bm k)$ favors the odd-parity superconductivity~\cite{Yoshida2012}.  
On the other hand, in the 2H$_a$ structure, a considerable interlayer hopping integral appears 
in the entire Brillouin zone, and therefore, we will see rather conventional behaviors in Nb-based TMDCs. 
Second, the magnetic field parallel to the conducting plane favors the odd-parity superconductivity while avoiding 
the orbital depairing effect, although the perpendicular field assumed in the multilayer Rashba systems\cite{Yoshida2012,Yoshida2014,Higashi2016} 
drastically suppresses the superconductivity. 
Fortunately, the Ising superconductivity in TMDCs is robust against the parallel magnetic field~\cite{Saito2015,Lu2015,Xi2015}.

The 2D superconductors in parallel magnetic fields have been investigated in the context of 
the Fulde-Ferrell-Larkin-Ovchinnikov (FFLO) state~\cite{FF,LO,Matsuda_review,Buzdin_review} 
and the helical superconducting state~\cite{NCSC,Agterberg_review}. 
However, we show that the odd-parity PDW state is more stable than 
the FFLO and helical states because of the peculiar symmetry of bilayer TMDCs.

This paper is organized as follows. 
In Sec.~II, the model for bilayer TMDCs is introduced, and the mean field theory is explained. 
We show the main results in Secs.~III and IV. In Sec.~III, the superconducting phase diagram in the Pauli limit 
is shown. In Sec.~IV, we calculate the phase diagram by taking into account both paramagnetic and orbital effects 
of the parallel magnetic field. It is shown that the odd-parity superconducting state, 
called the PDW state, is stable in intercalated bilayer 2H$_b$-TMDCs. 
In Sec.~V, the roles of the Rashba-type ASOC are examined. 
In Sec.~VI, a brief summary is given, and characteristic properties of the odd-parity PDW state are discussed 
for a future experimental test.

\section{Formulation}

\subsection{Bilayer model}

We investigate superconductivity in bilayer TMDCs by taking intrinsic Zeeman-type ASOC into account. 
We focus on the electron-doped TMDCs in which superconductivity has been reported~\cite{Ye2012,Costanzo2016,Jo2015,Shi2015,Saito2015,Lu2015,Xi2015,Woollam1977}, and we adopt 
a single-orbital tight-binding model for the transition metal $d_{z^2}$-orbital~\cite{Xiao_review}, 
\begin{align}
{\hat H}&={\hat H}_{0}+{\hat H}_{\rm I}.
\label{(1)}
\end{align}
The single-particle Hamiltonian is composed of 
\begin{align}
{\hat H}_{0}&={\hat H}_{\rm k}+{\hat H}_{\perp}+{\hat H}_{\rm Z}+{\hat H}_{\rm R} + {\hat H}_{\rm P}, 
\end{align}
where 
\begin{align}
{\hat H}_{\rm k}&=\sum_{{\bf k},m,s}\varepsilon({\bf k}+{\bf p}_{m}) \, c^{\dagger}_{{\bf k} m s}c_{{\bf k} m s}, \\
{\hat H}_{\perp}&=t_{\perp} \sum_{{\bf k}, s} f_{\perp}({\bf k}) \, c^{\dagger}_{{\bf k} 1 s}c_{{\bf k} 2 s} + h.c., \\
{\hat H}_{\rm Z}&=\sum_{{\bf k}, m ,s,s'}\alpha_{\rm Z}^{(m)}
\mbox{\boldmath ${\it g}$}_{\rm Z}({\bf k}+{\bf p}_{m})\cdot\mbox{\boldmath $\sigma$}_{s s'} \,
c^{\dagger}_{{\bf k} m s}c_{{\bf k} m s'}, \\
{\hat H}_{\rm R}&=\sum_{{\bf k},m,s,s'}\alpha_{\rm R}^{(m)}
\mbox{\boldmath ${\it g}$}_{\rm R}({\bf k}+{\bf p}_{m})\cdot\mbox{\boldmath $\sigma$}_{s s'} \,
c^{\dagger}_{{\bf k} m s}c_{{\bf k} m s'}, \\
{\hat H}_{\rm P}&=-\frac{g\mu_{\rm B}}{2}\sum_{{\bf k}, m, s, s'} {\bf H} \cdot \mbox{\boldmath $\sigma$}_{s s'} \,
c^{\dagger}_{{\bf k} m s}c_{{\bf k} m s'}, 
\end{align}
with $c^{\dagger}_{{\bf k} m s}$ being the creation operator for electrons with momentum ${\bf k}$ and spin $s$ on 
the $m$-th layer. The index for the layer takes $m=1,2$ in bilayer systems.

The first term is the kinetic energy term by hopping integrals in the 2D plane. 
Thus, we have 
\begin{align}
\varepsilon({\bf k})&=2t_{\rm 1}\left(\cos k_{\rm y}a+2\cos \frac{\sqrt{3}}{2}k_{\rm x}a \cos\frac{1}{2}k_{\rm y}a\right)-\mu,
\end{align}
by taking into account the nearest-neighbor hopping in the triangular lattice. 
The chemical potential $\mu$ is included in the dispersion relation $\varepsilon({\bf k})$. 
We fix 2D carrier density per layer $n_{\rm 2D} = 1 \times 10^{14}$ cm$^{-2}$ throughout this paper. 
This carrier density is close to the optimal doping of superconducting MoS$_2$~\cite{Ye2012}, 
and then, small Fermi surfaces enclose the $K$ and $K'$ points in the Brillouin zone.  
Later, the carrier density dependence is discussed. 
The lattice constant is assumed to be $a=3.2$\AA $\,\,$ in accordance with first principles 
calculations~\cite{Brumme2015,Fang2015} and an experimental report~\cite{Podberezskaya2001}. 
We choose the unit of energy $t_1 =1$, which is estimated to be $t_1 \simeq 200$ meV~\cite{Liu2013}.

 The second term ${\hat H}_{\perp}$ is the interlayer hopping energy which depends on the stacking structure.  
The interlayer hybridyzation function is 
\begin{align}
f_{\perp}({\bf k}) & = 1, 
\end{align} 
for the 2H$_a$ structure, while it is 
\begin{align}
f_{\perp}({\bf k}) &= \frac{1}{3} \Bigl[\cos\frac{k_{\rm x}}{\sqrt{3}}a+i\sin\frac{k_{\rm x}}{\sqrt{3}}a
\nonumber \\
&+2\left(\cos\frac{k_{\rm x}}{2\sqrt{3}}a-i\sin\frac{k_{\rm x}}{2\sqrt{3}}a\right)\cos\frac{1}{2}k_{\rm y}a \Bigr],
\label{f_perp}
\end{align}
for the 2H$_b$ structure. 
For non-intercalated TMDCs, we assume $t_\perp/t_1 =0.6$ in accordance with the band structure calculation 
for the bilayer MoS$_2$~\cite{Brumme2015}. However, much smaller $t_\perp$ is adopted for intercalated TMDCs, 
since it has been shown that the interlayer hopping is significantly decreased by the intercalation~\cite{Mattheiss}.

 The third and fourth terms represent ASOCs. 
The structure of ASOC has been classified by group theory~\cite{Frigeri_thesis}. 
Although the Rashba-type ASOC in polar point groups has been studied intensively~\cite{NCSC}, 
other kinds of ASOC may appear in non-polar point groups. 
Indeed, 21 point groups out of a total of 32 are noncentrosymmetric. 
Interestingly, the 2D materials classified into the $D_{\rm 3h}$ point group show a uniaxial ASOC, 
which causes spin polarization along the crystallographic {\it c}-axis. Such ASOC called ``Zeeman-type'' 
ASOC~\cite{Saito2015} is represented by ${\hat H}_{\rm Z}$, in which the spin texture in the momentum space is 
given by 
\begin{align}
\mbox{\boldmath ${\it g}$}_{\rm Z}({\bf k})&=
\frac{2}{3\sqrt{3}}\left(0, 0, \sin k_{\rm y}-2\cos\frac{\sqrt{3}}{2}k_{\rm x}\sin\frac{1}{2}k_{\rm y}\right). 
\label{g-vector_Z}
\end{align}
This term arises from the intrinsic (local) inversion symmetry breaking in the crystal structure of trigonal prismatic TMDCs.
We also take into account the Rashba-type ASOC term ${\hat H}_{\rm R}$, which comes from the bilayer 
structure~\cite{Maruyama2012}. 
Taking into account the nearest-neighbour coupling, we have the g-vector 
\begin{align}
\mbox{\boldmath ${\it g}$}_{\rm R}({\bf k})&=
\frac{1}{1.7602}\Bigl(-\sin k_{\rm y}-\cos \frac{\sqrt{3}}{2}k_{\rm x}\sin \frac{1}{2}k_{\rm y}, 
\nonumber \\
& \hspace{18mm} \sqrt{3}\sin \frac{\sqrt{3}}{2}k_{\rm x}\cos \frac{1}{2}k_{\rm y}, 0\Bigr). 
\label{g-vector_R}
\end{align}
The constant factors are chosen so that the maximum amplitude is unity, 
that is, ${\rm Max}_{\bf k} \, |f_{\perp}({\bf k})| = 
{\rm Max}_{\bf k} \, |\mbox{\boldmath ${\it g}$}_{\rm Z}({\bf k})| = 
{\rm Max}_{\bf k} \, |\mbox{\boldmath ${\it g}$}_{\rm R}({\bf k})| = 1$. 
Because the global inversion symmetry is recovered by the bilayer stacking, the layer-dependent coupling constants 
change sign, $(\alpha_{\rm Z}^{(1)}, \alpha_{\rm Z}^{(2)}) = (\alpha_{\rm Z}, -\alpha_{\rm Z})$ and 
$(\alpha_{\rm R}^{(1)}, \alpha_{\rm R}^{(2)}) = (\alpha_{\rm R}, -\alpha_{\rm R})$, and thus the spatial averages vanish. 
This is the sublattice-dependent ASOC which is characteristic of locally noncentrosymmetric systems~\cite{Fischer2011,Maruyama2012}. 
A coupling constant of Zeeman-type ASOC, $\alpha_{\rm Z}/t_1 =0.0375$, is adopted unless mentioned otherwise. 
Then, the spin splitting energy 
on the Fermi surface is $2 \alpha_{\rm Z}|\mbox{\boldmath ${\it g}$}_{\rm Z}({\bf k}_{\rm F})| \simeq 13$meV 
in accordance with the first principles band structure calculation for MoS$_2$~\cite{Saito2015}.  
The band structure calculation has also shown that the Rashba-type ASOC is much smaller than the Zeeman-type 
one~\cite{Saito2015}. Thus, we set $\alpha_{\rm R}=0$ except in Sec.~V.

 In this paper, we study superconducting states in a parallel magnetic field, whose effects appear in two ways. 
 One of the effects of the magnetic field is the paramagnetic effect, which is represented by the Zeeman coupling term 
${\hat H}_{\rm P}$. This term plays an essential role in this work. 
For simplicity, we adopt the g-factor $g=2$ and fix the direction of the magnetic field along the [100]-axis. 
 The other effect is the orbital effect taken into account through the Peierls phase. 
When the magnetic field is parallel to the 2D plane, the Peierls phase leads to a shift in momentum 
${\bf k} \rightarrow {\bf k} + \frac{e}{\hbar} {\bf A}$. For the magnetic field along the [100]-axis, 
we can choose the vector potential ${\bf A} = -H z {\hat y}$. Then, Eqs.~(3), (5), and (6) are modified 
by ${\bf p}_m = (3/2 -m) \frac{e}{\hbar} H c \, {\hat y}$ with $c$ being the lattice constant along the $c$-axis. 
We set $c=6.15$\AA $\,\,$for non-intercalated TMDCs~\cite{Schutte1987,Boker2001,Ramasubramaniam2011} 
and assume $c=1.5 \times 6.15$\AA $\,\,$$=9.225$\AA $\,\,$for intercalated TMDCs~\cite{Woollam1977}. 
Thus, the orbital effect is enhanced by intercalation.  

Many previous works have theoretically studied superconducting states in the paramagnetic field. 
It has been shown that the FFLO state is stable in a high magnetic field of centrosymmetric 
systems~\cite{Matsuda_review,Buzdin_review}, while the helical state is stabilized in a low magnetic field of 
noncentrosymmetric systems~\cite{NCSC,Agterberg_review}. 
Contrary to those studies, we show that the odd-parity PDW state may be more stable in bilayer TMDCs than 
the FFLO and helical states because of the local noncentrosymmetricity in the crystal structure.

 In order to study superconductivity emerging from the peculiar electronic state, a momentum-independent 
$s$-wave pairing interaction is introduced, 
\begin{align}
{\hat H}_{\rm I}&=-V\sum_{i, m} n_{i m \uparrow} n_{i m \downarrow}, 
\end{align}
with $n_{i m s}$ being the number density operator for spin $s$ at the site $i$ on the layer $m$. 
Although the Zeeman-type and Rashba-type ASOCs induce spin-triplet $p$-wave and $f$-wave components 
in the order parameter, respectively, we assume a purely $s$-wave pairing interaction for simplicity. 
This is justified because the effects of small spin-triplet components on the superconducting phase diagram 
are negligible~\cite{Yoshida2014}. 
We do not address the possibility of non-$s$-wave superconductivity by strong electron correlations~\cite{Yuan-Law2014,Hsu2016,Yuan-Honerkamp2015,Roldan2013}. 
First principles calculations for the $s$-wave superconductivity by electron-phonon 
interactions~\cite{Ge2013,Rosner2014,Das2015} reproduce the observed superconducting dome~\cite{Ye2012}, 
supporting our assumption. 
We fix the transition temperature of superconductivity at zero magnetic field, $T_{\rm c0}=5$K, by tuning the 
coupling constant $V$.

\subsection{Mean field theory}

 We analyze the model by means of the mean field theory. 
 When the superconducting state is spatially uniform in the 2D plane, the layer-dependent order parameter 
$\Delta_m = V \sum \langle c_{{\bf k} m \uparrow} c_{-{\bf k} m \downarrow} \rangle$ 
is calculated by decoupling the interaction term, 
\begin{align}
\hat{H}_{\rm I} \rightarrow
\sum_{{\bf k},m}\left(\Delta_{m}c^{\dagger}_{{\bf k} m \uparrow} c^{\dagger}_{{\bf -k} m \downarrow}+h.c.\right)+\sum_{m}\frac{\left|\Delta_{m}\right|^2}{V}.
\end{align}
Then, the Bogoliubov-de Gennes (BdG) Hamiltonian is diagonalized by transforming the basis 
\begin{align}
\hat{C}^{\dagger} &= \left( c^{\dagger}_{{\bf k} 1 \uparrow},c^{\dagger}_{{\bf k} 1 \downarrow},\cdots,
c_{{\bf -k} 2 \uparrow},c_{{\bf -k} 2 \downarrow} \right)
\nonumber \\
& \rightarrow \left({\hat \gamma}^{\dagger}_{1 {\bf k}},\cdots,{\hat \gamma}^{\dagger}_{8 {\bf k}} \right) 
= \hat{C}^{\dagger} U({\bf k}), 
\end{align}
with the unitary matrix, 
\begin{align}
 U({\bf k})=
  \begin{pmatrix}
   u_{1 \uparrow 1}({\bf k}) &\cdots &u_{1 \uparrow 8}({\bf k})\\
   \ddots &\vdots  &\ddots\\
   v_{2 \downarrow 1}({\bf k}) &\cdots &v_{2 \downarrow 8}({\bf k})\\
  \end{pmatrix}. 
\end{align}
The stationary solution for the layer-dependent order parameter is obtained by the self-consistent equation, 
\begin{align}
\Delta_{m}=-V \sum_{{\bf k}, \nu}u_{m \uparrow \nu}({\bf k})v^{*}_{m \downarrow \nu}({\bf k})f[E_{\nu}({\bf k})], 
\end{align}
where $E_{\nu}({\bf k})$ is the $\nu$-th eigenenergy at the momentum ${\bf k}$. 
The free energy is obtained by calculating 
\begin{align}
 F= & \frac{1}{2}\sum_{{\bf k},\nu}E_{\nu}({\bf k})f[E_{\nu}({\bf k})]
 +\frac{1}{2}T\sum_{{\bf k},\nu}f[E_{\nu}({\bf k})]\ln f[E_{\nu}({\bf k})]
\nonumber \\
& +\sum_{m}\frac{\left|\Delta_{m}\right|^2}{V}. 
\end{align}
The thermodynamically stable state is determined by comparing the free energy of metastable states. 
The BCS state corresponds to the solution $\Delta_1 = \Delta_2$, and the PDW state 
is characterized by the sign changing order parameter, $\Delta_1 = -\Delta_2$. 
The odd-parity superconductivity is realized in the PDW state by the sublattice degree of freedom, 
without requiring the spin-triplet pairing. 
We confirmed that the metastable solutions are spatially uniform in the absence of the orbital effect 
and Rashba-type ASOC. 
In Secs.~IV and V, we discuss the roles of the orbital effect and the Rashba-type ASOC, 
respectively. However, in the main part of this paper (Sec.~III), these minor effects are neglected.

 The numerical calculation for spatially non-uniform states requires a long computational time 
owing to long coherence length resulting from the small energy scale of superconductivity, $T_{\rm c0} \simeq 5$K. 
Thus, we solve the linearized BdG equation instead of solving the full BdG equation, when we study the 
non-uniform superconducting state in Secs.~IV and V. 
Superconducting states near the second order critical point are elucidated by the linearized theory 
capturing the divergence of superconducting susceptibility, ${\hat \chi}_{m m'}({\bf q})$. 
Adopting the T-matrix approximation, we obtain  
\begin{align}
{\hat \chi}({\bf q})=\frac{{\hat \chi}^0({\bf q})}{{\hat 1}-V {\hat \chi}^0({\bf q})}, 
\end{align}
where the irreducible susceptibility is calculated by 
\begin{align}
 {\hat \chi}^{0}_{m m'}({\bf q})&=
\nonumber \\
k_{\rm B}T\sum_{{\bf k},{l}} &\left[G^{\uparrow \uparrow}_{m m'}({\bf q/2}+{\bf k},i\omega_{l})
G^{\downarrow \downarrow}_{m m'}({\bf q/2}-{\bf k},-i\omega_{l})\right. \nonumber \\
& -\left.G^{\uparrow \downarrow}_{m m'}({\bf q/2}+{\bf k},i\omega_{l})
G^{\uparrow \downarrow}_{m m'}({\bf q/2}-{\bf k},-i\omega_{l})\right].
\end{align}
$G^{s s'}_{m m'}({\bf k},i\omega_{l})$ is the Green function on the Matsubara frequency, $\omega_{l}=(2l+1)\pi k_{\rm B}T$. 
The superconducting instability occurs when the maximum eigenvalue of $V {\hat \chi}_{}^0({\bf q})$ is unity. 
The eigenvector is proportional to 
$\Delta_m({\bf q}) = V \sum \langle c_{{\bf q/2}+{\bf k} m \uparrow} c_{{\bf q/2}-{\bf k} m \downarrow} \rangle$.

\begin{table}[htbp]
{\renewcommand\arraystretch{1.2}
\caption{Classification of superconducting states. Uniform BCS and PDW states as well as non-uniform FFLO, helical, Josephson vortex, and CS states are specified by the spatial dependence of order parameter (right column).}
\begin{center}
\begin{tabular}{c|c|c}
Uniform & BCS ($A_{\rm 1g}$) & $\Delta_1({\bf r}) = \Delta_2({\bf r}) = \Delta$  \\  \cline{2-3} 
(${\bf q}=0$) & PDW ($A_{\rm 2u}$) & $\Delta_1({\bf r}) = - \Delta_2({\bf r}) =\Delta$  \\ \hline
& LO & $\Delta_1({\bf r})  = \Delta_2({\bf r}) = \Delta \cos({\bf q} \cdot {\bf r})$ \\ \cline{2-3}
& FF/helical & $\Delta_1({\bf r})  = \Delta_2({\bf r})  = \Delta e^{i{\bf q} \cdot {\bf r}}$ \\ \cline{2-3}
Non-Uniform & Josephson vortex & $\Delta_1({\bf r}) = \Delta (e^{-i{\bf q} \cdot {\bf r}}+\delta e^{i{\bf q} \cdot {\bf r}})$ \\ 
(${\bf q} \ne 0$) &    & $\Delta_2({\bf r}) = \Delta (\delta e^{-i{\bf q} \cdot {\bf r}}+e^{i{\bf q} \cdot {\bf r}})$  \\  \cline{2-3} 
&CS & $\Delta_1({\bf r}) = \Delta (e^{i{\bf q} \cdot {\bf r}}+\delta e^{-i{\bf q} \cdot {\bf r}})$ \\ 
&    & $\Delta_2({\bf r}) = \Delta (\delta e^{i{\bf q} \cdot {\bf r}}+e^{-i{\bf q} \cdot {\bf r}})$ \\
\end{tabular}
\label{tab1}
\end{center}
}
\end{table} 

The classification of superconducting states is summarized in Table~\ref{tab1}. 
The uniform superconducting states are BCS and PDW states. The concept of PDW state was introduced for spatially 
inhomogeneous superconducting states in the atomic length scale which is much shorter than the coherence length~\cite{Agterberg_PDW}. 
In the subsequent works~\cite{Yoshida2012,Yoshida2014,Higashi2016,Sumita2016,Yoshida2015,Watanabe2015,Yoshida_Tsuneya2016}, however, the sign changing order parameter between sublattices has also been classified 
into the PDW state. 
In the latter case, the translation symmetry is not broken (${\bf q}=0$). 
On the other hand, the irreducible representation of order parameter differs from the BCS state. 
While the BCS state belongs to the $A_{\rm 1g}$ irreducible representation, the PDW state belongs to the 
odd-parity $A_{\rm 2u}$ irreducible representation of $D_{\rm 3d}$ point group.

When the center-of-mass momentum ${\bf q}$ is finite, the translation symmetry is broken. 
Then, the superconducting state may be classified into 
the FFLO~\cite{FF,LO,Matsuda_review,Buzdin_review}, helical~\cite{NCSC,Agterberg_review}, 
complex-stripe (CS)~\cite{Yoshida2013}, or Josephson vortex states. 
The detailed definition of these non-uniform states has been given in Ref.~\onlinecite{Watanabe2015}. 
In the case studied here, the Josephson vortex state (Secs.~IV and V) and CS state (Sec.~V) may be stable. 
Both states are regarded as an intermediate state between the Fulde-Ferrell (FF) state~\cite{FF} 
and the Larkin-Ovchinnikov (LO) state~\cite{LO}. The order parameter is a superposition of 
$\Delta_m({\bf q})$ and $\Delta_m(-{\bf q})$. Then, both amplitude and phase of 
order parameter are spatially non-uniform~\cite{Yoshida2013}, although only the amplitude (phase) is 
non-uniform in the LO (FF) state.

\section{Phase diagram in the Pauli limit}

\begin{figure}[htbp]
 \begin{center}
\includegraphics[keepaspectratio, width=9.0cm]{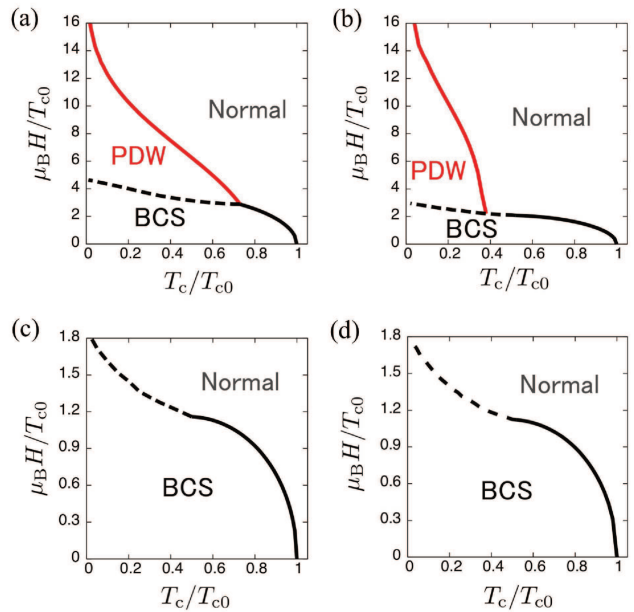}
  \caption{(Color online) 
Superconducting phase diagram in the Pauli limit. 
(a-c) 2H$_b$ stacking structure with (a) $t_\perp/t_1 = 0.06$, (b) $t_\perp/t_1 = 0.125$, and (c) $t_\perp/t_1 = 0.6$. 
(d) 2H$_a$ stacking structure with $t_\perp/t_1 = 0.125$. 
Solid (dashed) line shows the second order (first order) transition. 
The PDW state may be stable in the high magnetic field region of the 2H$_b$ structure, 
although the BCS state is stable in the whole superconducting phase of the 2H$_a$ structure. 
We assume $\alpha_{\rm Z}=0.0375$ in accordance with MoS$_2$. 
}
  \label{fig:non-vortex_diagram}
 \end{center}
\end{figure}

First, we examine the superconducting phase diagram in the Pauli limit by setting ${\bf p}_m=0$. 
Figure~\ref{fig:non-vortex_diagram} shows the $H$-$T$ phase diagram of 2H$_b$ structure. 
For small interlayer hopping integrals, $t_\perp/t_1 = 0.06$ and $t_\perp/t_1 = 0.125$, the odd-parity PDW state 
is stable in the high magnetic field region [Figs.~\ref{fig:non-vortex_diagram}(a) and (b)]. 
Although the PDW state is not stable for a moderate interlayer hopping, $t_\perp/t_1 = 0.6$, consistent with 
non-intercalated MoS$_2$ [Fig.~\ref{fig:non-vortex_diagram}(c)], $t_\perp/t_1$ is decreased in the intercalated 
TMDCs~\cite{Mattheiss}, suggested here as a candidate of odd-parity superconductors.

\begin{figure}[htbp]
\begin{center}
\includegraphics[keepaspectratio, width=8.5cm]{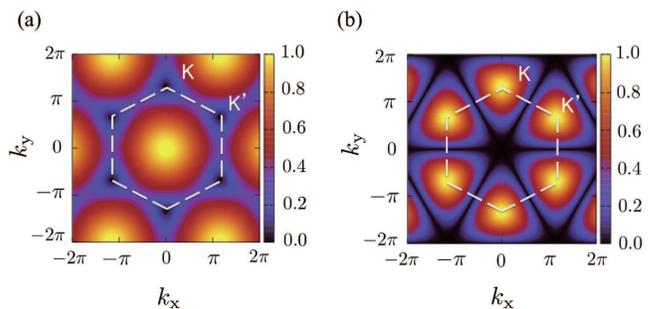}
\caption{(Color online) 
Momentum dependence of (a) the interlayer hybridyzation function $|f_\perp({\bf k})|$ for the 2H$_b$ structure and 
(b) the magnitude of Zeeman-type ASOC, $|\mbox{\boldmath ${\it g}$}_{\rm Z}({\bf k})|$. 
Hexagonal white dashed line shows the Brillouin zone boundary, whose vertex is the $K$ and $K'$ points. 
} 
\label{fig:BZ}
\end{center}
\end{figure}

The 2H$_b$ stacking structure plays an important role in stabilizing the PDW state. 
Indeed, the 2H$_a$ structure realizes only the conventional BCS state [Fig.~\ref{fig:non-vortex_diagram}(d)] 
even when the interlayer hopping is small. 
The distinct difference between the 2H$_a$ and 2H$_b$ structures comes from the interlayer hybridyzation function, 
$f_{\perp}({\bf k})$. Although it is momentum-independent in the 2H$_a$ structure, {\it the $f_{\perp}({\bf k})$ disappears 
at the $K$ and $K'$ points in the 2H$_b$ structure} because of the quantum interference of three interlayer 
hopping integrals~\cite{Xiao_review,Akashi2014,Akashi2016}. 
Figure~\ref{fig:BZ} plots the momentum dependence of the interlayer hybridyzation function, (\ref{f_perp}), 
as well as the magnitude of the Zeeman-type g-vector, (\ref{g-vector_Z}). 
We see that $f_{\perp}({\bf K})=f_{\perp}({\bf K'})=0$ while the g-vector $\mbox{\boldmath ${\it g}$}_{\rm Z}({\bf k})$ 
takes the maximum amplitude at ${\bf k}={\bf K}$ and ${\bf K}'$. 
Therefore, the ratio of the ASOC and interlayer hybridyzation diverges at the $K$ point, 
\begin{align}
\alpha_{\rm Z} |\mbox{\boldmath ${\it g}$}_{\rm Z}({\bf K})|/t_\perp |f_\perp({\bf K})| = \infty. 
\label{ratio_K}
\end{align}
Since the Fermi momentum is in the vicinity of the $K$ or $K'$ point, the ratio may be large, 
\begin{align}
\alpha_{\rm Z} |\mbox{\boldmath ${\it g}$}_{\rm Z}({\bf k}_{\rm F})|/t_\perp |f_\perp({\bf k}_{\rm F})| > 1, 
\label{ratio_kf}
\end{align}
on the Fermi surface even when $\alpha_{\rm Z}/t_\perp \ll 1$. 
The disappearance of the interlayer hybridyzation is protected by the 3-fold rotation 
symmetries~\cite{Xiao_review,Akashi2014,Akashi2016}, and thus above features are not an artifact 
of our model but general properties of the 2H$_b$ structure without orbital degeneracy. 

Previous studies have shown that the sublattice-dependent ASOC plays an important role when the ratio 
$\alpha_{\rm Z} |\mbox{\boldmath ${\it g}$}_{\rm Z}({\bf k}_{\rm F})|/t_\perp |f_\perp({\bf k}_{\rm F})|$ 
is large~\cite{Maruyama2012}. 
For the PDW state to be stabilized, the condition 
$\alpha_{\rm Z} |\mbox{\boldmath ${\it g}$}_{\rm Z}({\bf k}_{\rm F})|/t_\perp |f_\perp({\bf k}_{\rm F})| \geq 1 $ 
has to be satisfied~\cite{Yoshida2012}.
Thus, the relation (\ref{ratio_kf}) indicates the stable PDW state in the 2H$_b$ structure. 
For the parameters $\alpha_{\rm Z}=0.0375$, $t_\perp =0.125$ and $n_{\rm 2D} = 1 \times 10^{14}$ cm$^{-2}$, 
the ratio is $\alpha_{\rm Z} |\mbox{\boldmath ${\it g}$}_{\rm Z}({\bf k}_{\rm F})|/t_\perp |f_\perp({\bf k}_{\rm F})| =1.68> 1$, 
satisfying the condition.  
A smaller carrier density makes Fermi momentum to be closer to the $K$ point, and then the ratio 
$\alpha_{\rm Z} |\mbox{\boldmath ${\it g}$}_{\rm Z}({\bf k}_{\rm F})|/t_\perp |f_\perp({\bf k}_{\rm F})|$ is increased.
Therefore, the PDW state is furthermore stable in the low carrier density region.  

On the other hand, the condition for the thermodynamically stable PDW state is approximately reduced to 
$\alpha_{\rm Z}/t_\perp \geq 1$ in the 2H$_a$ structure since $f_\perp({\bf k})=1$. 
This condition is hardly satisfied in NbSe$_2$ having a moderate spin-orbit coupling. 
Therefore, the odd-parity superconductivity is unlikely in the bilayer 2H$_a$-NbSe$_2$~\cite{Xi2015}.

\begin{figure}[htbp]
 \begin{center}
\includegraphics[keepaspectratio, width=9.0cm]{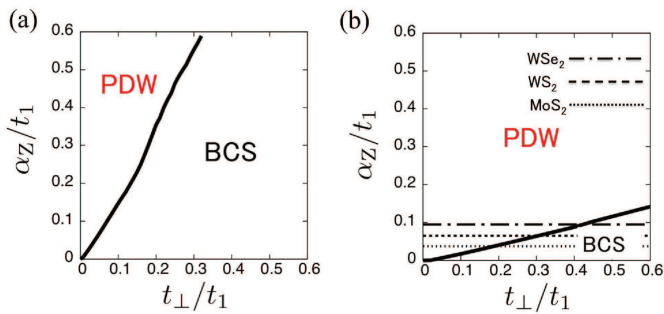}
  \caption{(Color online)
Phase diagram as a function of $\alpha_{\rm Z}$ and $t_\perp$ for (a) the 2H$_a$ structure and (b) the 2H$_b$ structure. 
The temperature is fixed to be $T/T_{\rm c0}=0.1$ and the magnetic field is tuned so as to be just below 
the upper critical field. In (b) the dotted, dashed, and dot-dashed lines show $\alpha_{\rm Z}/t_1$ 
in MoS$_2$, WS$_2$, and WSe$_2$, respectively. 
}
  \label{fig:non-vortex_diagram2}
 \end{center}
\end{figure}

The contrasting behavior of the 2H$_a$ and 2H$_b$ structures is illuminated by Fig.~\ref{fig:non-vortex_diagram2}, 
which shows the phase diagram in the $\alpha_{\rm Z} \,$-$\, t_\perp$ plane at a low temperature $T/T_{\rm c0}=0.1$. 
In the 2H$_a$ structure, the PDW state is stable only when  $\alpha_{\rm Z}/t_\perp \geq 2$.
On the other hand, the condition for the PDW state is significantly 
relaxed in the 2H$_b$ structure to $\alpha_{\rm Z}/t_\perp \geq 0.25$.

\section{orbital effect}

Next, we examine the orbital effect on the superconducting phases. 
Although the orbital effect is completely suppressed in the monolayer, it may affect bilayer TMDCs. 
Since we consider the magnetic field parallel to the 2D plane, the Abrikosov vortex lattice state is not realized. 
However, the orbital effect may induce the Josephson vortex which penetrates into the atomic bilayers. 

The Josephson vortex state is characterized by an order parameter with finite 
center-of-mass momentum ${\bf q}$~\cite{Watanabe2015}, and thus it is distinguished from the PDW state 
as well as from the BCS state. 
We calculate the upper critical field for various center-of-mass momentum ${\bf q}$ of Cooper pairs by solving  
the linearized BdG equation. 
The superconducting state with the highest upper critical field is stable near the critical point.

\begin{figure}[htbp]
\begin{center}
\includegraphics[keepaspectratio, width=9cm]{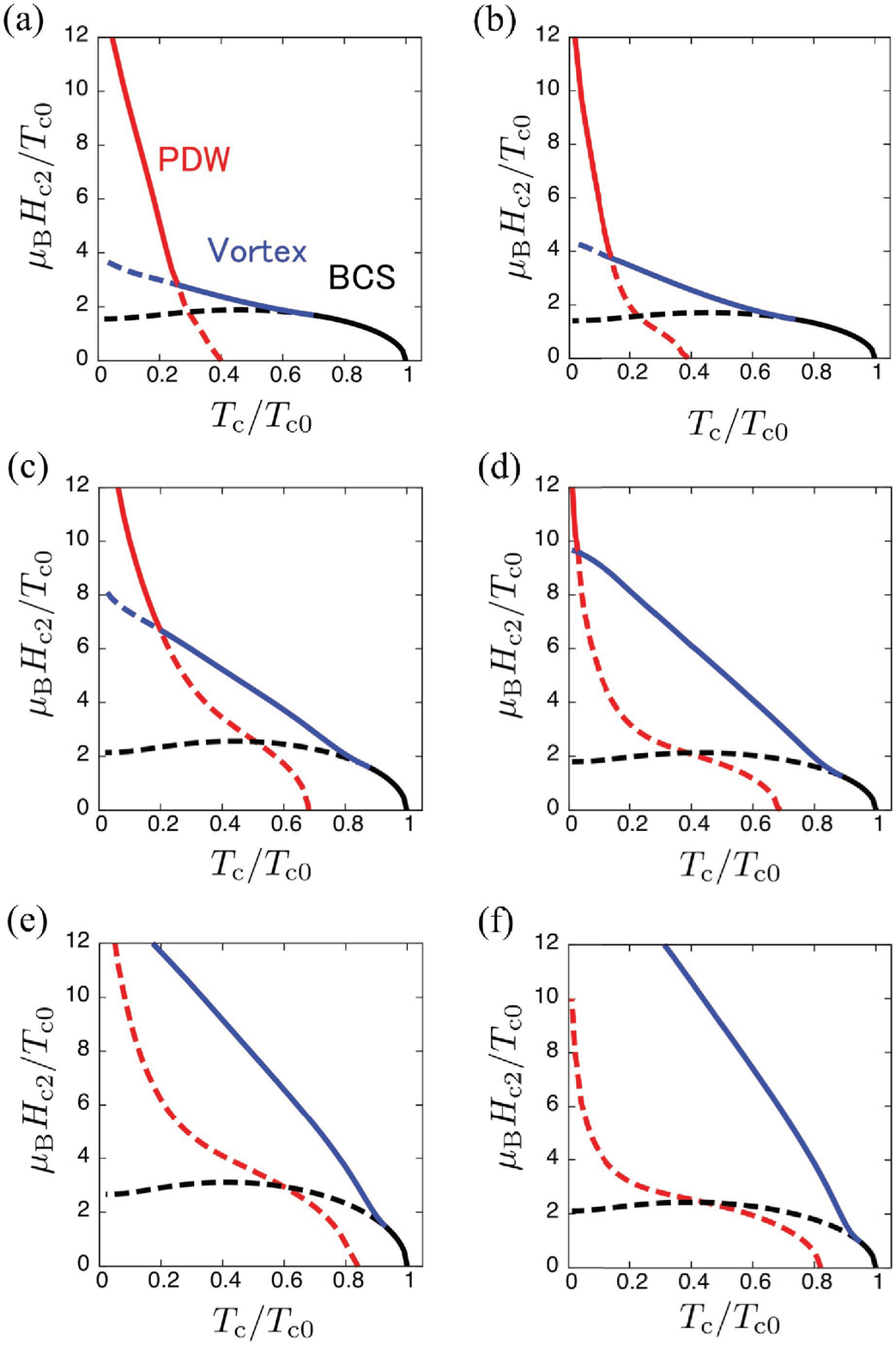}
\caption{(Color online) 
Upper critical fields of the BCS (black), PDW (red), and Josephson vortex (blue) states. 
The Zeeman-type ASOC is chosen to be (a,b) $\alpha_{\rm Z} = 0.0375$, (c,d) $\alpha_{\rm Z} = 0.065$, 
and (e,f) $\alpha_{\rm Z} = 0.095$, in agreement with MoS$_2$, WS$_2$, and WSe$_2$, respectively. 
The other parameters are the same as Fig.~\ref{fig:non-vortex_diagram}(b). 
The highest (realizable) upper critical field is shown by the solid line, while the fictitious ones 
are drawn by the dashed lines. 
The $c$-axis lattice constant is set to (a,c,e) $c=6.15$\AA $\,\,$for non-intercalated TMDCs and 
(b,d,f) $c=9.225$\AA $\,\,$for intercalated TMDCs, respectively.
} 
\label{fig:vortex_diagram}
\end{center}
\end{figure}

Figures~\ref{fig:vortex_diagram}(a) and (b) show the upper critical fields of the BCS, PDW, and Josephson vortex states 
for the same parameters as Fig.~\ref{fig:non-vortex_diagram}(b). The increase in the $c$-axis lattice constant by 
intercalation~\cite{Woollam1977} is taken into account in Fig.~\ref{fig:vortex_diagram}(b), although it is neglected 
in  Fig.~\ref{fig:vortex_diagram}(a). 
Both figures show that the PDW state in the 2H$_b$ structure is robust against the orbital effect 
at low temperatures although the Josephson vortex state is stable in the intermediate temperature region.

On the other hand, we find that the PDW state is suppressed when the Zeeman-type ASOC is furthermore increased. 
Although the parameters compatible with intercalated MoS$_2$ have been assumed in Fig.~\ref{fig:vortex_diagram}(b), 
the ASOC is increased in Figs.~\ref{fig:vortex_diagram}(d) and (f) while maintaining the other parameters. 
The coupling constants $\alpha_{\rm Z}=0.065$ and $\alpha_{\rm Z}=0.095$ are consistent with the spin splitting energy 
at the $K$ point $\sim 26$meV and $\sim 38$meV in WS$_2$ and WSe$_2$, respectively. 
Then, we obtain the ratio, 
$\alpha_{\rm Z} |\mbox{\boldmath ${\it g}$}_{\rm Z}({\bf k}_{\rm F})|/t_\perp f_\perp({\bf k}_{\rm F}) = 2.91$ and $4.25$, respectively. 
As the Zeeman-type ASOC is increased, the upper critical field of the Josephson vortex state is enhanced 
although that of the PDW state is insensitive to the ASOC. 
Therefore, the PDW state becomes less stable than the Josephson vortex state. 
For a large ASOC compatible with WSe$_2$, the PDW state is completely suppressed [Fig.~\ref{fig:vortex_diagram}(f)]. 
Thus, we conclude that the intercalated bilayer MoS$_2$ is the best platform for the odd-parity PDW state 
rather than the heavier TMDCs.

The ASOC dependence is understood by paying attention to the 
paramagnetic depairing effect. 
The PDW state completely avoids the paramagnetic depairing effect, 
because the symmetry of superconductivity is the same as spin triplet superconductivity~\cite{Maruyama2012}. 
On the other hand, the BCS and Josephson vortex states are suppressed by the paramagnetic depairing effect. 
This is indeed the reason why the PDW state is stable in the high magnetic field region~\cite{Yoshida2012}.
However, the Zeeman-type ASOC protects the BCS and Josephson vortex states against the paramagnetic depairing effect. 
Hence, the upper critical field of the Josephson vortex state increases by increasing the ASOC. 
These features have been demonstrated in Fig.~\ref{fig:vortex_diagram}. 
The Josephson vortex state has also been demonstrated in a recent study for bilayer TMDCs~\cite{Liu2016}, 
although the PDW state has not been shown. 
Combining with the results in the paramagnetic limit (Sec.~III), we find that an intermediate value of the Zeeman-type ASOC 
satisfying, 
\begin{align}
1 \leq \alpha_{\rm Z} |\mbox{\boldmath ${\it g}$}_{\rm Z}({\bf k}_{\rm F})|/t_\perp |f_\perp({\bf k}_{\rm F})| \leq 3,
\end{align} 
favors the odd-parity PDW state. 
The condition may actually be satisfied in the intercalated MoS$_2$ and WS$_2$ as well as in the 
non-intercalated TMDCs with heavy metal ions and/or small carrier density. 
As we mentioned previously, the ratio 
$\alpha_{\rm Z} |\mbox{\boldmath ${\it g}$}_{\rm Z}({\bf k}_{\rm F})|/t_\perp |f_\perp({\bf k}_{\rm F})|$ can be tuned 
by the carrier density, for which the electrostatic control has been demonstrated 
in various TMDCs~\cite{Ye2012,Saito2015,Costanzo2016,Jo2015,Shi2015}.

\begin{figure}[htbp]
\begin{center}
\includegraphics[keepaspectratio, width=9.0cm]{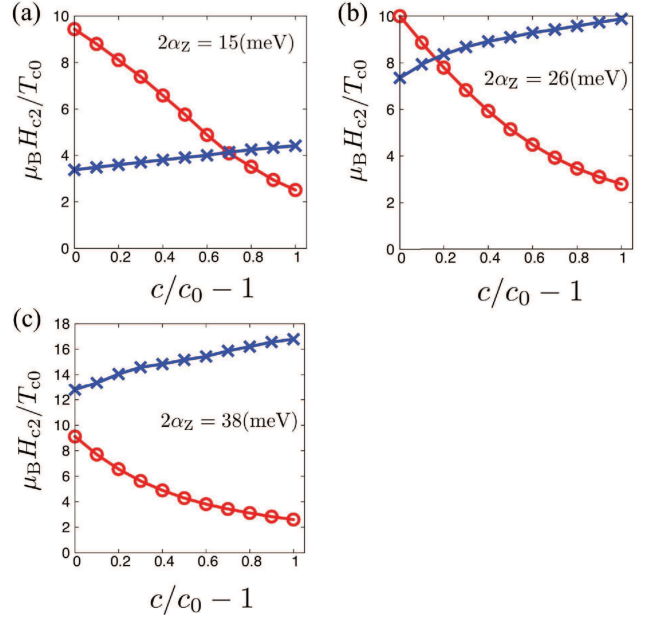}
\caption{(Color online) 
Upper critical fields of the PDW state (red circles) and Josephson vortex state (blue crosses) at $T/T_{\rm c0}=0.1$ 
as a function of the interlayer distance normalized by $c_0 =6.15$\AA $\,$ of non-intercalated MoS$_2$. 
(a) $\alpha_{\rm Z} = 0.0375$, (b) $\alpha_{\rm Z} = 0.065$, and (c) $\alpha_{\rm Z} = 0.095$. 
The other parameters are the same as Fig.~\ref{fig:vortex_diagram}. 
} 
\label{fig:c-dependence}
\end{center}
\end{figure}

At the end of this section, we discuss the effect of increased interlayer distance by intercalation. 
The superconductivity has been observed in intercalated K$_x$MoS$_2$, Rb$_x$MoS$_2$, and Cs$_x$MoS$_2$ 
which maintain the hexagonal crystal structure~\cite{Woollam1977}. 
The $c$-axis lattice constant increases by 35\%-60\%. The two effects by intercalation have been taken into account: 
the interlayer hopping integral $t_\perp$ is decreased, and the orbital effect is enhanced. 
We illustrate here the enhanced orbital effect by Fig.~\ref{fig:c-dependence}, which shows the lattice 
constant dependence of the upper critical fields for a fixed  $t_\perp$. 
It is revealed that the PDW state (Josephson vortex state) is suppressed (slightly enhanced) 
by the orbital effect as the interlayer distance is increased. 
However, the PDW state is still stable for the parameters, $\alpha_{\rm Z} = 0.0375$ and $c=9.225$\AA, 
compatible with the intercalated MoS$_2$.

\section{Rashba spin-orbit coupling}

Up to now we have ignored the Rashba-type ASOC, because it is negligible in MoS$_2$. 
Even in a strong external electric field the Rashba-type ASOC is less than 2\% of the Zeeman-type ASOC~\cite{Saito2015}. 
Such a small Rashba term does not alter the superconducting phase diagram. 
However, we discuss here an alternative way to realize the odd-parity PDW state using a large Rashba-type ASOC, 
considering the tunability of spin-orbit coupling by heterostructure engineering.

\begin{figure}[htbp]
\begin{center}
\includegraphics[keepaspectratio, width=9.0cm]{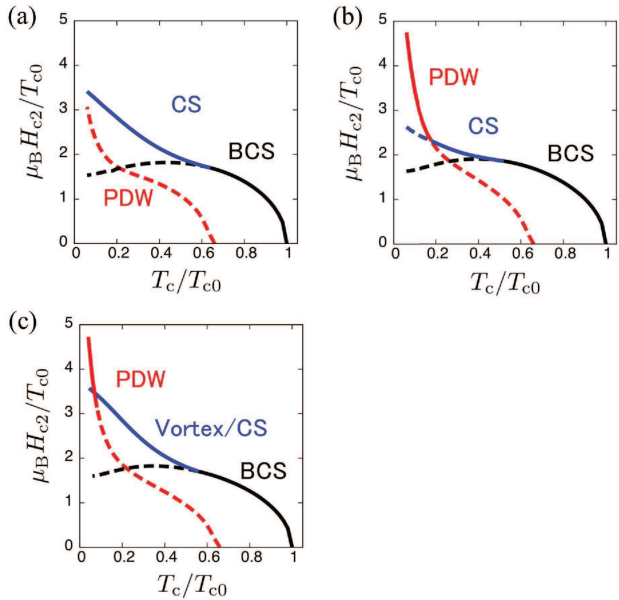}
\caption{(Color online) 
Upper critical fields of the BCS (black), PDW (red), and Josephson vortex/CS (blue) states 
in the presence of the Rashba-type ASOC. (a) $\alpha_{\rm R} = \pm 0.185$ and the orbital effect is neglected (Pauli limit). 
Taking the orbital effect into account, we obtain 
(b) for $\alpha_{\rm R} = + 0.185$ and (c) for $\alpha_{\rm R} = - 0.185$. 
The other parameters are $\alpha_{\rm Z} = 0.0375$, $t_\perp/t_1 =0.125$, and $c=6.15$\AA.
} 
\label{fig:Rashba-type_ASOC}
\end{center}
\end{figure}

Figure~\ref{fig:Rashba-type_ASOC} shows the phase diagram for the Rashba-type ASOC, $\alpha_{\rm R}= \pm 0.185$. 
Then, the Rashba-type ASOC is comparable to the Zeeman-type ASOC on the Fermi surface, 
$|\alpha_{\rm R} \mbox{\boldmath ${\it g}$}_{\rm R}({\bf k}_{\rm F})| \simeq 
|\alpha_{\rm Z} \mbox{\boldmath ${\it g}$}_{\rm Z}({\bf k}_{\rm F})|$, for $\alpha_{\rm Z} = 0.0375$. 
The phase diagram in the Pauli limit [Fig.~\ref{fig:Rashba-type_ASOC}(a)] does not show the stable PDW state, 
in contrast to Fig.~\ref{fig:non-vortex_diagram}(b), which shares the other parameters. 
Thus, the Rashba-type ASOC suppresses the PDW state in the Pauli limit. 
Instead, the CS state with finite Cooper pairs' momentum (see Table~\ref{tab1}) 
is stable in the high magnetic field region. 
The obtained CS state is essentially the same as that obtained in the bilayer Rashba model~\cite{Yoshida2013}.

On the other hand, the combination of the Rashba-type ASOC and the orbital effect stabilizes the PDW state.
The direction of the Cooper pairs' center-of-mass momentum ${\bf q}$ in the CS state is determined by the sign of 
the Rashba-type ASOC. In our model, the ${\bf q}$ is opposite between the CS state and the Josephson vortex state 
when $\alpha_{\rm R} > 0$. 
Then, the effects of Rashba-type ASOC and the orbital effect are canceled, and therefore, the PDW state is stable 
[Fig.~\ref{fig:Rashba-type_ASOC}(b)], as at $\alpha_{\rm R} =0$ in the Pauli limit. 
Even in the opposite case, $\alpha_{\rm R} < 0$, the PDW state is stable in a small parameter range  
[Fig.~\ref{fig:Rashba-type_ASOC}(c)]. 
These results are qualitatively consistent with what we observed in the bilayer Rashba model~\cite{Watanabe2015}.

\section{Summary and discussion}

In this work, we show the odd-parity superconductivity in bilayer TMDCs with 2H$_b$ stacking structure. 
Under the parallel magnetic field, the Zeeman-type ASOC arising from the intrinsic local inversion symmetry breaking 
realizes the $\pi$-junction of the spin-singlet $s$-wave order parameter between two atomic layers. 
Such a non-uniform superconducting state in the atomic scale is called the PDW state~\cite{Yoshida2012}. 
The sign changing order parameter belongs to the $A_{2u}$ irreducible representation of the $D_{3d}$ point group. 
Thus, the parity of superconductivity is odd. 
Although the odd-parity superconductivity has attracted great interests for more than four decades, only a few 
spin-triplet superconductors such as Sr$_2$RuO$_4$~\cite{Maeno2012}, UPt$_3$~\cite{Joynt2002}, and ferromagnetic 
superconductors~\cite{Saxena2000,Aoki2012} have been identified as candidate materials. 
The conditions favoring the spin-triplet Cooper pairing are hardly satisfied in natural materials. 
Our finding shows a new and realizable mechanism of odd-parity superconductivity by means of the symmetry control 
enabled by the van der Waals heterostructure. 
Tuning the conventional $s$-wave superconductors by the spin-orbit coupling makes the odd-parity superconductivity.

 The bilayer 2H$_b$-TMDCs are promising platform of the odd-parity superconductivity, 
because two important conditions are naturally satisfied. First, the magnetic field required to stabilize 
the PDW state is parallel to the 2D plane, and therefore, the orbital effect suppressing the superconductivity is 
substantially avoided. 
Second, the interlayer hybridization vanishes at the $K$ point in the Brillouin zone owing to the quantum interference. 
Then, the Zeeman-type ASOC overcomes the interlayer hybridization which suppresses the PDW state. 
Calculating the superconducting phase diagram by taking into account both paramagnetic and orbital effects, 
we conclude that the intercalated bilayer MoS$_2$ and WS$_2$ are a platform for the odd-parity superconductivity. 
Recently, superconductivity has been realized in the bilayer MoS$_2$ by gating~\cite{Costanzo2016}.

Finally, we discuss some properties of the PDW state to be examined by future experiments. 
(1) The superconducting gap in the density of states is increased~\cite{Yoshida2012}. 
(2) The spin susceptibility is enhanced~\cite{Maruyama2012}. 
(3) The vortex core radius shrinks~\cite{Higashi2016}. 
(4) The zero energy Andreev bound states appear at vortex cores in the tilted magnetic field~\cite{Higashi2016}. 
(5) The upper critical field exceeds the Pauli-Chandrasekhar-Clogston limit and shows upward curvature at low temperatures~\cite{Yoshida2012}. 
The features of the upper critical field have been universally observed in the intercalated hexagonal bulk 
MoS$_2$~\cite{Woollam1977,Woollam1976}, indicating a signature of the spin-orbit coupling in the superconducting state. 
In contrast, the upper critical field of non-hexagonal TMDCs does not exceed the Pauli-Chandrasekhar-Clogston 
limit~\cite{Woollam1976}. This systematic behavior points to our view on the importance of the 2H$_{b}$ crystal structure. 

The features (1) and (2) can be tested by thermodynamic or spectroscopic experiments, and (3) and (4) may be detected 
by measurements in real space, such as scanning tunneling microscopy (STM) and nuclear magnetic resonance (NMR). 
Although these experiments may be challenging, developments in the technology of artificial 2D electron 
systems may clarify the exotic superconducting properties. 
The realization and identification of odd-parity superconductivity in a controllable way would be 
a great development in the research field of superconductivity with exotic symmetry and topology.

\section*{Acknowledgements} 
The authors are grateful to R. Akashi, Y. Iwasa, K. T. Law, T. Nojima, Y. Saito, 
T. Yoshida, N. F. Q. Yuan, and T. Watanabe for fruitful discussions. 
Y. N. is supported by a JSPS Fellowship for Young Scientists. 
This work was supported by Grant-in Aid for Scientific Research on Innovative Areas ``J-Physics'' (Grant No. JP15H05884) 
and ``Topological Materials Science'' (Grant No. JP16H00991) from JSPS of Japan, and by JSPS KAKENHI Grant Numbers  
JP15K05164 and JP15H05745.

\end{document}